\documentclass[useAMS,usenatbib,fleqn]{mn2e}
\usepackage{graphicx}
\usepackage{textcomp}
\usepackage{amsmath}
\usepackage{breqn}
\usepackage{wasysym}
\usepackage{amssymb}
\title[Formalism for exoplanet transmission spectroscopy]{An analytical formalism 
accounting for clouds and other ``surfaces'' for exoplanet transmission spectroscopy}
\author[Yan B\'etr\'emieux and Mark R. Swain]{Yan B\'etr\'emieux$^{1}$ 
\thanks{NASA Postdoctoral Program Fellow}
\thanks{E-mail: Yan.Y.Betremieux@jpl.nasa.gov}
and Mark R. Swain$^{1}$\\
$^{1}$Jet Propulsion Laboratory, California Institute of Technology, 4800 Oak Grove Drive,
Pasadena, CA 91109, USA}
\begin{document}


\pagerange{\pageref{firstpage}--\pageref{lastpage}} \pubyear{2016}

\maketitle

\label{firstpage}

\begin{abstract}

Although the formalism of \citet{Lecavelier_2008} is extremely useful to understand what shapes transmission spectra of exoplanets, it does not include the effects of a sharp change in flux with altitude generally associated with surfaces and optically thick clouds. Recent advances in understanding the effects of refraction in exoplanet transmission spectra have, however, demonstrated that even clear thick atmospheres have such a sharp change in flux due to a refractive boundary. We derive a more widely applicable analytical formalism by including first-order effects from all these ``surfaces" to compute an exoplanet's effective radius, effective atmospheric thickness, and spectral modulation for an atmosphere with a constant scale height. We show that the effective radius cannot be located below these ``surfaces" and that our formalism matches \citet{Lecavelier_2008}'s in the case of a clear atmosphere. Our formalism explains why clouds and refraction reduce the contrast of spectral features, and why refraction decreases the Rayleigh scattering slope as wavelength increases, but also shows that these are common effects of all ``surfaces". We introduce the concept of a ``surface" cross section, the minimum mean cross section that can be observed, as an index to characterize the location of ``surfaces" and provide a simple method to estimate their effects on the spectral modulation of homogeneous atmospheres. We finally devise a numerical recipe that extends our formalism to atmospheres with a non-constant scale height and arbitrary sources of opacity, a potentially necessary step to interpret observations.

\end{abstract}


\section{Introduction}\label{intro}
With the upcoming launches of the Transiting Exoplanet Survey Satellite (TESS) and the James Webb Space Telescope (JWST), the amount and quality of exoplanetary transmission spectra available for interpretation are expected to increase dramatically. To analyze the anticipated wealth of data in a timely and statistically meaningful fashion, many members of the exoplanet community combine sophisticated statistical algorithms (\citealt{Irwin_2008}; \citealt{M_S_2009}; \citealt{B_S_2012}; \citealt*{Lee_2012}; \citealt{Line_2013a}; \citealt{Gibson_2014}; \citealt{Waldmann_2015}) with instrument models and radiative transfer codes to retrieve atmospheric properties of transiting exoplanets from their lightcurves (see e.g. analyses by \citealt{Barstow_2013}; \citealt{Line_2013b}; \citealt*{Swain_2014}; \citealt{Griffith_2014}; \citealt{McCullough_2014}; \citealt{Kreidberg_2014b}, \citealt{Dragomir_2015}; \citealt{Kreidberg_2015}; \citealt{Tsiaras_2016}; \citealt{Evans_2016}; \citealt{Iyer_2016}) and to plan observational strategies for JWST (\citealt{Deming_2009}; \citealt{Barstow_2015}; \citealt{Greene_2016}). As many of these algorithms must compute numerous exoplanetary spectra as they sample across the allowed multi-dimensional space of atmospheric parameters, the forward models which compute these spectra must strike a proper balance between speed and accuracy.

In light of this, analytical formalisms to compute exoplanetary spectra are extremely useful. They allow one to understand how key atmospheric parameters shape exoplanetary spectra, can constitute the backbone of forward models, and increase computational speed tremendously if they are general and accurate enough. They can also test, under the same conditions for which they are valid, the accuracy of more general numerical methods. One important formalism for transmission spectroscopy, against which radiative transfer codes are sometimes compared (see e. g. \citealt{Shabram_2011}; \citealt{H_B_2012}) and which a few analyses use (see e.g. \citealt{Dragomir_2015}; \citealt{Sing_2015}, 2016) is the semi-analytical derivation of \citet{Lecavelier_2008} for a homogeneous and isothermal atmosphere. 

From numerical simulations, \citet{Lecavelier_2008} show that for a broad range of planetary radius to scale height ratios ($R_P/H$), the effective radius ($R_{eff}$) of a transiting exoplanet is located at an altitude where the integrated optical depth (or slant optical depth, $\tau$) along a ray grazing the planetary limb is about 0.56. They build upon this numerical result to show analytically that the variation with wavelength of the effective radius of an exoplanet is proportional  to the atmospheric scale height. They also demonstrate that in the presence of an opacity source which obeys a simple known power law with wavelength, such as Rayleigh scattering by gases or well-mixed hazes, one can deduce the scale height of the atmosphere by measuring the spectral slope of the Rayleigh signature (Rayleigh slope), provided it is the only source of opacity in the observed spectral region.

Later, \citet{dW_S_2013} confirm in a purely analytical fashion \citet{Lecavelier_2008}'s results and also show that the relevant scale height of the atmosphere changes with the type of the dominant source of opacity, with collision-induced absorption (CIA) having half the effective scale height of Rayleigh scattering. 
The analytical derivation of \citet{dW_S_2013} is extremely important because it explains under which conditions the formalism of \citet{Lecavelier_2008} holds, namely:
\begin{itemize}
\item{Atmospheric scale height ($H$) is constant with altitude ($z$).}
\item{Atmospheric composition is homogeneous so that the mean extinction cross section of the atmosphere is also constant with altitude.}
\end{itemize}

\begin{figure}
\includegraphics[scale=0.55]{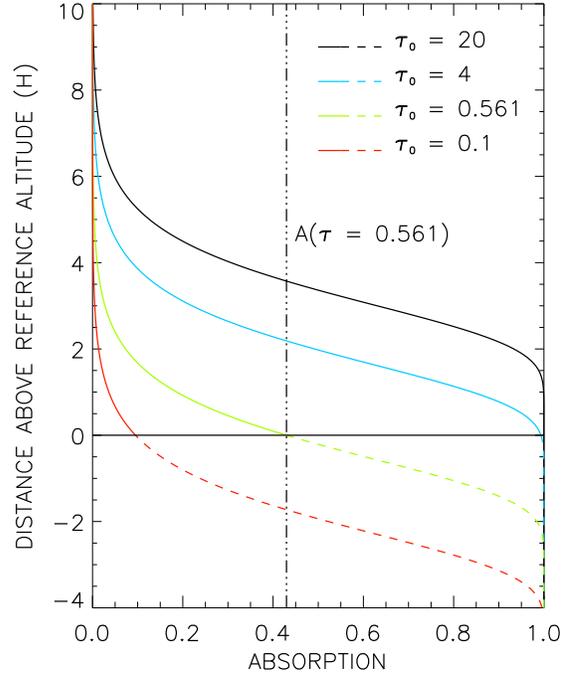}
\caption{Limb absorption profile of a clear atmosphere as a function of distance above a reference altitude (in units of scale height) for different values of the slant optical depth of the atmosphere ($\tau_0$) at this reference altitude. According to \citet{Lecavelier_2008} and \citet{dW_S_2013}, the effective radius of an exoplanet is located at an altitude where the optical depth of a ray traversing the atmosphere is about 0.561 (i.e. where the vertical triple-dot-dashed line intersects the various limb absorption profiles). If the reference altitude is the location of a ``surface", regions below the ``surface" are completely opaque and the absorption profile increases to 1, thus differing from the absorption profile of a clear atmosphere (dashed lines) in these regions.} \label{fig1}
\end{figure}

A direct consequence of these two conditions is that the limb absorption profile ($A$) of the atmosphere is fully described as a function of the distance ($\Delta z$) above a reference altitude by specifying the slant optical depth ($\tau_0$) along a ray grazing this reference altitude. The limb absorption profile is indeed then given by
\begin{equation}
A \equiv 1 - e^{-\tau} = 1 - e^{-\tau_0 e^{-\Delta z/H}} \label{absorbprofile}
\end{equation}
which Fig.~\ref{fig1} depicts as a function of distance in scale height ($\Delta z/H$) above this reference altitude for different values of $\tau_0$. Changing the value of $\tau_0$ does not modify the shape of the absorption profile but merely shifts it in altitude. Since the effective radius of an exoplanet only depends on its limb absorption profile \citep{Brown_2001}, the effective radius is always located at the same point in this particular profile (i.e. where $\tau = 0.561$) as demonstrated by \citet{dW_S_2013}.

Thus, both formalisms are very useful for clear isothermal homogeneous atmospheres, or ones where the hazes are well-mixed, because the limb absorption profile follows closely that in equation~\ref{absorbprofile}. However, as illustrated in Fig.~\ref{fig1}, a planetary surface or an optically thick cloud deck creates a sharp boundary (or ``surface") below which stellar radiation is cut-off, which can distort the limb absorption profile sufficiently that the formalism of \citet{Lecavelier_2008} breaks down. When $\tau_0 = 20$ the ``surface" has no impact on the effective radius of an exoplanet because the atmosphere is completely opaque at the ``surface" and the ``surface" does not effectively distort the limb absorption profile. A slant optical depth of 0.561 \citep{dW_S_2013} is thus representative of the location of the effective radius of the exoplanet for this purely atmospheric profile. When $\tau_0 = 4$, the limb absorption profile deviates from that in equation~\ref{fig1} because it increases sharply to one at the ``surface". As $\tau_0$ decreases further, the difference in absorption between the ``surface" and the atmosphere increases, and the limb absorption profile deviates increasingly from that in equation~\ref{fig1} (shown by dashed lines below the ``surface"). When $\tau_0 < 0.561$, the absorption profile is sufficiently distorted that \citet{Lecavelier_2008}'s formalism completely breaks down and predicts that the effective radius of the exoplanet is below the ``surface". However, common sense dictates that it should be above, as there is still some atmospheric absorption. Indeed, for an airless rocky exoplanet where atmospheric absorption is non-existent, its effective radius is simply located at its surface.

So not only are these formalisms (\citealt{Lecavelier_2008}; \citealt{dW_S_2013}) not appropriate for terrestrial exoplanets with optically thin atmospheres, but neither are they for atmospheres where a cloud deck mimics a surface. Unfortunately, there is a growing body of evidence that many exoplanets may have clouds or hazes. GJ1214b is the most compelling example, as its transmission spectrum remains featureless to the best photometric precision, a fact which is more likely explained by a high altitude cloud deck than a high mean molecular weight atmosphere \citep{Kreidberg_2014a}. Indeed, cloud decks decrease the contrast of spectral features (e. g. \citealt{B_S_2012}; \citealt{H_B_2012}; \citealt{B_S_2013}; \citealt{YB_LK_2014}) and are the likely cause for the reduced contrast, compared with that expected from a clear atmosphere, of the 1.4~$\mu$m water band (\citealt{Sing_2016}; \citealt{Stevenson_2016}; \citealt{Iyer_2016}), sodium and potassium lines (\citealt{Sing_2016}; \citealt{Heng_2016}) observed in a few hot Jupiters. 

Even thick cloudless atmospheres, or one with well-mixed hazes, require an analysis incorporating a ``surface" as recent results (\citealt{Sidi_Sari_2010}; \citealt{YB_LK_2015}; \citealt{YB_2016}) show that thick atmospheres have a sharp change in flux due to refraction. Indeed, a refracted distorted image of the stellar surface is superimposed on the limb absorption profile \citep{Munoz_Mills_2012}, and the refractive boundary can be predominantly due to a change in the back-illumination of the atmosphere at the imaged stellar limb, unlike clouds and surfaces which only change the limb absorption profile. The spectroscopic signature of this first-order refractive effect and an optically thick cloud deck are similar in nature because they both cause lower atmospheric regions to appear dark in transmission spectroscopy (see Fig.~2 in \citealt{YB_LK_2014}). The major difference is the altitude where they occur, where for example the refractive boundary of an Earth-Sun analog is located above tropospheric clouds, significantly reducing the contrast of water features in the visible and near infrared (\citealt{Munoz_2012}; \citealt{YB_LK_2013}, 2014; \citealt*{Misra_2014}). Moreover, the location of a refractive boundary is easier to compute than that of a cloud top because the chemical species which contribute to an atmosphere's refractive properties are the same that contribute to its mean molecular weight, whereas the chemical nature of clouds are currently difficult to retrieve from the lightcurve of a transiting exoplanet. \citet{YB_2016} further shows that refraction modifies the Rayleigh slope, and that it can break the degeneracy between retrieved abundances of chemical species and the planet's radius.

Nevertheless, a few recent papers use \citet{Lecavelier_2008}'s formalism as a starting point for their own analytical derivations. \citet{Line_Parmentier_2016} explore the effects of a non-homogeneous limb distribution of vertically well-mixed hazes on an exoplanet's transmission spectrum. \citet{Heng_2016} derives an index quantifying the haziness of an atmosphere from the contrast between the core and wings of sodium and potassium lines. Clearly, a new formalism which can also account for ``surfaces" (surfaces, optically-thick cloud decks, or refractive boundaries) is badly needed by the community.

In this paper, we derive the effective radius and the effective atmospheric thickness of a transiting exoplanet by proceeding along the line of \citet{dW_S_2013}'s derivation. We highlight where our results differ from theirs, and show that we find an extra term which accounts for the presence of a ``surface". We explore the various ramifications in terms of the location of the exoplanet's effective radius, its change with wavelength, and its effects on the Rayleigh slope. We discuss how ``surfaces" decrease the contrast of spectral features, and devise a figure of merit for characterizing the location of a ``surface" and a formalism for quickly estimating its impact on the spectral modulation of an atmosphere. We also show the differences that the choice of formalism makes on the computed transmission spectrum. Finally, we develop a numerical recipe to generalize our formalism to atmospheres with a non-constant scale height and arbitrary sources of opacity.

\begin{figure}
\includegraphics[scale=0.27]{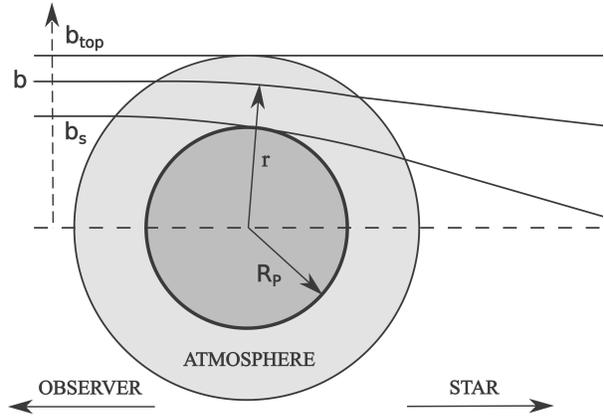}
\caption{Geometry of observation of a transiting exoplanet. When the center of the star is occulted, this picture is cylindrically symmetric about the horizontal dashed line. Rays traversing the atmosphere and reaching the observer are confined to impact parameters ($b$) between $b_s$ and $b_{top}$. The above picture is for a refractive atmosphere which bends rays, where $r$ is the grazing radius (i.e. point of closest approach) of a ray. In the non-refractive approximation, rays follow straight lines parallel to the dashed line and $b = r$.
\label{newfig2}}
\end{figure}

\section{Analytical derivation}\label{derivation}

Our derivation for an exoplanet's effective radius is similar to the one in the main article and the Supplementary Materials of \citet{dW_S_2013}. It is accurate for an atmosphere with a constant scale height, ignoring stellar limb darkening and second-order refractive effects. We highlight differences in results, as well as differences in notation when necessary. 

Two quantities appear regularly throughout this paper: the impact parameter of a ray (or projected distance to the center of the exoplanet with respect to the observer, $b$), and the grazing radius of a ray (or closest approach to the center of the exoplanet, $r$). Both quantities are illustrated in Fig.~\ref{newfig2} which shows the geometry of observation of a transiting exoplanet. Even though we do not include the subtle secondary effects of refraction which creates differences between $b$ and $r$, we use the appropriate quantities in the various equations as if refraction was fully included. However, throughout the discussions in this paper, $b$ and $r$ are interchangeable because they are identical in our essentially non-refractive treatment. Nevertheless, one can include first-order ``surface" effects of refraction by locating the ``surface" at the impact parameter of the refractive critical boundary, whose location depends both on the properties of the atmosphere and the planet-star geometry \citep{YB_LK_2014}.

\subsection{Effective radius and atmospheric thickness}\label{effectiverad}

A transiting exoplanet causes a momentary drop ($\Delta F$) in the observed stellar flux ($F_\star$),
which is expressed in term of the effective radius ($R_{eff}$) of the exoplanet which occults the stellar surface by
\begin{equation}
 \left( \frac{R_{eff}}{R_{\star}} \right)^2 \equiv \frac{F_{out} - F_{in}}{F_{out}} \approx \frac{\Delta F}{F_\star}
\end{equation}
where $F_{out}$ and $F_{in}$ are the observed fluxes outside of and in the middle of the transit, respectively, and $R_{\star}$ is the stellar radius. Strictly speaking, $F_{out}$ is the sum of the stellar flux and
scattered radiation from a thin crescent-shaped region of the observable dayside of the exoplanet, as well as thermal emission from the nightside of the exoplanet. However, the exoplanet's contribution to the observed out-of-transit flux is negligible compared to the stellar flux \citep{Brown_2001}.

The effective radius of an exoplanet is the radius that a completely opaque spherical body occulting 
a uniformly bright star must have in order to duplicate the observed flux drop. Ignoring stellar limb darkening, the effective radius of an exoplanet occulting the center of its host star is theoretically computed by
\begin{equation}
R_{eff}^2 = b_s^2 + 2 \int_{b_s}^{b_{top}} (1 - e^{-\tau}) b db = b_s^2 + C \label{effradius}
\end{equation}
\citep{Brown_2001} where $\tau$ is the slant optical depth of a ray through the atmosphere, and $b_{s}$ and $b_{top}$ are the impact parameters of the ray at the ``surface" and the top of the atmosphere, respectively (shown in Fig.~\ref{newfig2}). The contribution to the effective radius (i.e. to the flux drop) is split between the completely opaque ``body" which lies below the ``surface" (first term), and that of the semi-transparent atmosphere above it (second term). We represent the atmospheric term by $C$  just like in \citet{dW_S_2013}, but replace their first term (i.e. planetary radius at the surface, $R_{p,0}$) with $b_s$. 

The effective atmospheric thickness ($h$) is the difference between the effective radius of the exoplanet and the ``surface", so that
\begin{equation}
R_{eff}^2 \equiv (b_s + h)^2 = b_s^2 + 2 b_s h + h^2 \label{effradius2} .
\end{equation}
Using equations~\ref{effradius} and~\ref{effradius2}, we can relate $C$ to the effective atmospheric thickness by
\begin{equation}
h^2 + 2 b_s h = C ,
\end{equation}
which has only one physical solution:
\begin{equation}
h = -b_s + \sqrt{b_s^2 + C} = b_s (-1 + \sqrt{1 + (C/b_s^2)}) \label{exacth}.
\end{equation}
We note the difference inside the square root between our expression and equation~6 in \citet{dW_S_2013}.
Their expression suffer from a typographical mistake, as evidenced by inconsistencies in the units on both sides of their equation. However, this does not affect their results.

Since $(C/b_s^2)$ is the ratio of the effective projected surfaces of the thin atmospheric annulus to that of the planetary body, we can usually expect that $(C/b_s^2) \ll 1$. In that case, we can rewrite equation~\ref{exacth} as a Taylor series and obtain
\begin{align}
h & \approx b_s \left[ -1 + (1 + \frac{1}{2} (C/b_s^2) - \frac{1}{8} (C/b_s^2)^2 + ... ) \right]  \notag \\
   & \approx \frac{C}{2 b_s} \left[ 1 - \frac{1}{4} (C/b_s^2) + ... \right] \label{effheight} .
\end{align}
The series inside the square bracket is a correction factor which is close to and always less than one.

\subsection{Slant optical depth}\label{opticaldepth}

The slant optical depth along the path of a ray depends on the nature of the opacity through the  relationship of the absorption coefficient with number density, as well as the vertical distribution of the chemical species responsible for the stellar radiation's extinction. Assuming a homogeneous atmosphere, such that the mole fraction ($f_i$) of each species is constant with altitude, and that the absorption coefficient is proportional to number density (e.g. Rayleigh scattering), the slant optical depth ($\tau_\sigma$) is given by
\begin{equation}
\tau_\sigma = N \sum_i f_i \sigma_i = N \sigma \label{opacity1}
\end{equation}
where $\sigma_i$ is the extinction cross section of each species, $\sigma$ is the mean atmospheric cross section, and $N$ ( $\equiv \left< N^1 \right>$ ) is the column-integrated density along the path of the ray (called horizontal integrated density by \citealt{Fortney_2005}), or simply column abundance, expressed in molecule cm$^{-2}$. In the case of collision-induced absorption (CIA), where the absorption coefficient is proportional to the square of the number density, the slant optical depth ($\tau_k$) is given by
\begin{equation}
\tau_k = \left< N^2 \right> \sum_{i,j} f_i f_j k_{i,j} = \left< N^2 \right> k \label{opacity2}
\end{equation}
where $k_{i,j}$ is the collision-induced cross-section of the $i^{th}$ and $j^{th}$ species, $k$ is the mean atmospheric collision-induced cross section, and $\left< N^2 \right>$ is the square of the density integrated along the path of the ray, expressed in molecule$^{2}$ cm$^{-5}$. We adopt the same units as HITRAN \citep{Richard_2012} for the collision-induced cross section, namely cm$^{5}$ molecule$^{-2}$.

In the case of an isothermal atmosphere, and assuming that the gravitational acceleration does not change much over the altitude regions which contribute significantly to an exoplanet's transmission spectrum, the neutral density ($n$) through the atmosphere is related to its value ($n_0$) at a reference radius ($r_0$) by 
\begin{equation}
n = n_0 e^{-(r - r_0)/H} \label{densdistr}
\end{equation}
where the density scale height of the atmosphere ($H$) is constant. Since the absorption coefficient of various sources of opacity can have different dependences with number density, let us consider the general case where the absorption coefficient is proportional to $n^q$. The integrated density to the $q^{th}$ power along the path of the ray ( $\left<N^{q} \right>$ ) is given by
\begin{equation}
\left<N^{q} \right>  = 2 \int_{0}^{\infty} n^q ds = 2 n_0^q \int_{0}^{\infty} e^{-q(r - r_0)/H} ds \label{colabun2} ,
\end{equation}
where $ds$ is a distance along the ray with respect to its grazing point. As already pointed out, $q = 1$ for Rayleigh scattering, and $q = 2$ for CIA. The situation for line transitions is more complicated as $q$ is equal to one near the core of the line and the far Lorentzian wings, and goes up to two in the Voigt-to-Doppler transition region of the line (see the comprehensive discussion in \citealt{dW_S_2013}'s Supplementary Materials). Since the pressure along the path of the ray changes by many orders of magnitude, the relative weight of Doppler and collisional broadening changes with altitude and so does the value of $q$ at a given wavelength. However, molecular bands are the sum of many overlapping individual transition lines, and we postulate that only the sides of bands, where the cross section is rapidly changing with wavelength, are dominated by this transition region. Hence, most spectral regions of a molecular cross section should fall outside this region, and into the domain where $q$ is about one. 

Provided no extra exponential term is hidden in $ds$, we can see by inspection that the exponential 
in equation~\ref{colabun2} is identical to that of a constant scale height atmosphere with a slant optical depth scale height ($H_\tau$), which we define by
\begin{equation}
H_\tau \equiv H/q   \label{sldepthscale} .
\end{equation}
That is indeed the case in a non-refractive treatment, as the exponential dependence with $H$ then only comes from the density dependence, in opposition to the general case which is fairly complex (see equations~32 and~17, respectively, in \citealt{YB_LK_2015}). Thus, the solution to equation~\ref{colabun2} in our non-refractive treatment is simply the leading term in equation~43 of \citet{YB_LK_2015} with $H_\tau$ replacing $H$. It is then 
\begin{equation}
\left<N^{q} \right>  =  \sqrt{2\pi r H_\tau} n_0^q e^{-(r - r_0)/H_\tau} \approx \left<N^{q} \right>_0 e^{-(r - r_0)/H_\tau}  \end{equation}
where we neglect the $r^{1/2}$ dependence ($r^{1/2} \approx r_0^{1/2}$) as the variation of $\left<N^{q} \right>$ with altitude is dominated by the exponential factor. $\left<N^{q} \right>_0$ is simply $\left<N^{q} \right>$ where the grazing radius is the reference radius. 

In general, the slant optical depth comes from a mixture of opacities with different $H_\tau$,  
\begin{equation}
\tau = \tau_\sigma + \tau_k ,
\end{equation}
and the slant optical depth scale height will be between those of Rayleigh scattering and CIA (i.e. $q$ is between one and two). Moreover, the relative importance of CIA opacities versus other opacities changes with altitude, so that $H_\tau$ might also change significantly with large changes in altitude. However, much of the discussions in the remainder of the paper assume that one form of opacity dominates in a given spectral region (i.e. $H_\tau$ is constant with altitude). We treat the general case in Section~\ref{integration}.

When $H_\tau$ is constant with altitude, it follows that 
\begin{equation}
\tau  = \tau_0 e^{-(r - r_0)/H_\tau}  \label{slantdepth}
\end{equation}
where $\tau$ is given by either equation~\ref{opacity1} or~\ref{opacity2} depending on the opacity source, and $\tau_0$ is the slant optical depth at the reference radius. When selecting the ``surface" as the reference radius, the very definitions of the effective slant optical depth ($\tau_{eff}$) and the effective atmospheric thickness ($h$) imply that they are related by
\begin{equation}
\tau_{eff}  \equiv \tau_s e^{-h/H_\tau}  \label{effslantdepth} ,
\end{equation}
where $\tau_s$ is the slant optical depth at the ``surface".

\subsection{Atmospheric contribution - integral form} \label{integral}

To determine the effective atmospheric thickness, we must evaluate
\begin{equation}
C = 2 \int_{b_s}^{b_{top}} (1 - e^{-\tau}) b db  .
\end{equation}
To solve this integral, we first use equation~\ref{slantdepth} with the ``surface" as the reference radius, then replace $r$ with $b$ because of our non-refractive treatment (see comments in Section~\ref{derivation}). We thus derive the relationships
\begin{equation}
b = b_s - H_\tau \ln(\tau/\tau_s) \label{varchange}
\end{equation}
and 
\begin{equation}
db = - H_\tau \frac{d\tau}{\tau} 
\end{equation}
which we use to change the integral variable from $b$ to $\tau$. This leads to 
\begin{equation}
C = 2 H_\tau b_s \int_\delta^{\tau_s} (1 - e^{-\tau}) \frac{d\tau}{\tau} - 2 H_\tau^2 \int_\delta^{\tau_s} (1 - e^{-\tau}) \ln(\tau/\tau_s) \frac{d\tau}{\tau} \label{intform}
\end{equation}
where $\delta \rightarrow 0$ as it is the slant optical depth at the top of the atmosphere. The second term is on the order of $(H_\tau/b_s)$ larger than the first term. Since $(H_\tau/b_s) \sim (H_\tau/R_p)$ is usually much smaller than one, we neglect the second term. This term probably contributes to the departure of the effective slant optical depth from a constant when $R_p/H$ is small (see Fig.~1 of \citealt{Lecavelier_2008}). However, we focus here on the domain which is applicable to the largest ensemble of exoplanets.

\subsection{Atmospheric contribution - solution} \label{solution}

Solving the integral in the first term of equation~\ref{intform} is a multi-step process. The first step consists in splitting the integral by changing the limits of integration 
\begin{equation}
I = \int_0^{\tau_s} (1 - e^{-\tau}) \frac{d\tau}{\tau} = \int_0^{1} (1 - e^{-\tau}) \frac{d\tau}{\tau} + \int_{1}^{\tau_s} (1 - e^{-\tau}) \frac{d\tau}{\tau} . \label{split}
\end{equation}
Equation~\ref{split} is true whether $\tau_s$ is larger or smaller than one. We then add and subtract identical integral terms to obtain
\begin{equation}
I = \gamma + \int_{1}^{\tau_s} (1 - e^{-\tau}) \frac{d\tau}{\tau} + \int_{1}^{\infty} e^{-\tau} \frac{d\tau}{\tau} \label{inter}
\end{equation}
where $\gamma$ ($\approx 0.577$) is the Euler-Mascheroni constant, whose definition can be found in \citet{Chandrasekhar} (by inspection from equations~9 and 10 in Appendix~1) and is
\begin{equation}
\gamma = \int_{0}^{1} (1 - e^{-\tau}) \frac{d\tau}{\tau} - \int_{1}^{\infty} e^{-\tau} \frac{d\tau}{\tau} .
\end{equation}
We then rewrite the integral terms in equation~\ref{inter} as
\begin{equation}
I = \gamma + \int_{1}^{\tau_s} \frac{d\tau}{\tau} + \int_{\tau_s}^{\infty} e^{-\tau} \frac{d\tau}{\tau} \label{complex} .
\end{equation}
The second term in equation~\ref{complex} is trivial and simply $\ln\tau_s$. The last term is the 
exponential integral $E_1(\tau_s)$ (see equation~7 in Appendix~1 of \citealt{Chandrasekhar}).
Hence, we get the very useful result that
\begin{equation}
I = \int_0^{\tau_s} (1 - e^{-\tau}) \frac{d\tau}{\tau} = \gamma + \ln\tau_s + E_1(\tau_s) \label{results} .
\end{equation}

Using this result in equation~\ref{intform}, and combining it with equations~\ref{effradius} or~\ref{effheight}, we can see that to first-order the effective radius of an exoplanet is given by
\begin{equation}
R_{eff}^2 \approx b_s^2 + 2 H_\tau b_s \left( \gamma + \ln\tau_s + E_1(\tau_s)  \right)
\end{equation}
and the effective atmospheric thickness by
\begin{equation}
h \approx \frac{C}{2 b_s} = H_\tau \left( \gamma + \ln\tau_s + E_1(\tau_s)  \right) \label{effatmthick} .
\end{equation}
We can compare our result for the effective atmospheric thickness to that in equation~7 of \citet{dW_S_2013}, noting that we use the symbol $\tau_s$ instead of $A_\lambda$, and $H_\tau$ instead of $R_{p,0}B$. We find our solutions are identical except that we have an extra term $H_\tau E_1(\tau_s)$. What is the physical significance or interpretation of this term?

\section{Ramifications}

\subsection{Significance of solution}

To understand the physical significance of our solution, we first rearrange the terms in equation~\ref{effatmthick} into
\begin{equation}
\ln\tau_s - h/H_\tau =  - \gamma - E_1(\tau_s) .
\end{equation}
We then take the exponential on both sides and use equation~\ref{effslantdepth} to get the very simple relation
\begin{equation}
\tau_{eff} = e^{-\gamma} e^{- E_1(\tau_s)} .
\end{equation}
It is interesting to note that $e^{-\gamma}$ ($\approx 0.561$) is the effective slant optical depth associated with \citet{dW_S_2013}'s solution which matches the simulations of \citet{Lecavelier_2008} for a clear atmosphere. However, our solution has an extra multiplicative factor which depends on the optical properties of the atmosphere at the ``surface". Does our solution matches their solution for a clear atmosphere? A clear atmosphere can be thought of as an atmosphere with a ``surface" buried so deeply that it can not be observed (i.e. $\tau_s \rightarrow \infty$). Since $E_1(\tau_s \rightarrow \infty) = 0$, our extra multiplicative factor is indeed one for a clear atmosphere.

\begin{figure}
\includegraphics[scale=0.55]{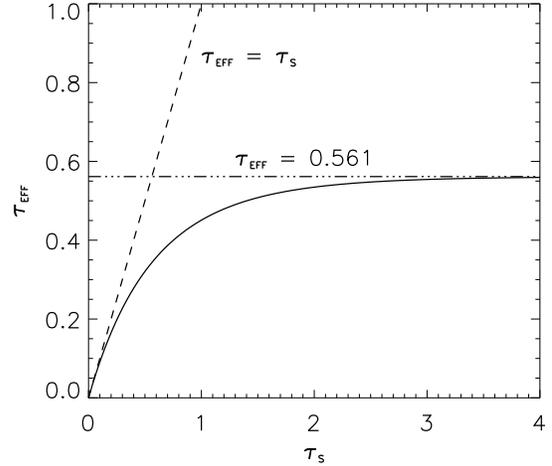}
\caption{Effective slant optical depth ($\tau_{eff}$, at the effective exoplanet radius) as a function of the slant optical depth of the atmosphere at the ``surface" ($\tau_s$). Also shown are the clear atmosphere solution of \citet{dW_S_2013} (triple-dot-dashed line) and when the effective radius is located at the ``surface" (dashed line).
\label{fig2}}
\end{figure}

Fig.~\ref{fig2} shows the behaviour of $\tau_{eff}$ with $\tau_s$ in its transition from the optically thick to the optically thin regime. As $\tau_s~\rightarrow~\infty$, $\tau_{eff}$ asymptotically approaches the clear atmosphere solution of \citet{dW_S_2013}. Below $\tau_s = 4$, $\tau_{eff}$ diverges from this solution as $\tau_s$ decreases, and asymptotically approaches the ``surface" described by the ``$\tau_{eff} = \tau_s$" dashed line. This makes sense because the effective radius is located at a slant optical depth within the bounds of the observable atmospheric region, i.e. above the ``surface". Thus, our extra integral exponential term from equation~\ref{effatmthick} accounts for the effect of a ``surface", and decreases the effective slant optical depth associated with the effective radius of the exoplanet. 

\subsection{Change of effective radius with wavelength}

How does this new result change the variation of the effective exoplanet radius with wavelength, originally discussed by \citet{Lecavelier_2008}? Assuming $H_\tau$ does not change much with wavelength,
\begin{equation}
\frac{dR_{eff}}{d\lambda} = \frac{dh}{d\lambda} \approx H_\tau \frac{d}{d\lambda} \left( \gamma + \ln\tau_s + E_1(\tau_s)  \right) .
\end{equation}
Differentiation of the exponential integral gives
\begin{equation}
\frac{dE_1(\tau_s)}{d\lambda} = \frac{-e^{-\tau_s}}{\tau_s} \left( \frac{d\tau_s}{d\lambda} \right) = -e^{-\tau_s} \frac{d(\ln\tau_s)}{d\lambda}
\end{equation}
so that
\begin{equation}
\frac{dR_{eff}}{d\lambda} \approx H_\tau \left( 1 - e^{-\tau_s} \right) \frac{d(\ln\tau_s)}{d\lambda}  .
\label{firstvarwave}
\end{equation}
As \citet{dW_S_2013}, we find that the size of spectral features depends on the nature of the opacity through $H_\tau$, with CIA spectral features proportional to half of the scale height (see equation~\ref{sldepthscale}). When the absorption coefficient is proportional to density, such as for Rayleigh scattering, this reduces to
\begin{equation}
\frac{dR_{eff,\sigma}}{d\lambda} \approx H \left( 1 - e^{-\tau_s} \right) \frac{d(\ln\sigma)}{d\lambda} \label{variancewave}
\end{equation}
because the cross section is the only quantity which is wavelength-dependent in the expression for the slant optical depth. For CIA, we simply replace $\sigma$ with $k$ (see equation~\ref{opacity2}), and $H$ with $H/2$ (see equation~\ref{sldepthscale}).

We find an extra multiplicative factor $(1 - e^{-\tau_s})$ compared to \citet{Lecavelier_2008}. This holds true also for differentiation with respect to $\ln\lambda$. For a clear atmosphere where $\tau_s$ tends to infinity, this extra factor tends to one. As $\tau_s$ tends to zero, this extra factor tends to $\tau_s$ and becomes vanishingly small. Hence, the size of spectral features is not strictly proportional to the scale height, but also depends on the slant optical thickness of the atmosphere at the ``surface",  which can become vanishingly small when the ``surface" number density is low and the observable atmosphere is optically thin. 

Similarly, this factor also modifies the Rayleigh slope in spectral regions where Rayleigh scattering dominates. As the Rayleigh scattering cross section decreases from  the ultraviolet to the infrared, so does its slant optical thickness. In spectral regions where the Rayleigh scattering slant optical thickness is less than about four, the Rayleigh slope decreases with wavelength until it becomes flat.

The decrease of the contrast of spectral features has been demonstrated several times for both clouds and refraction through radiative transfer numerical simulations of exoplanetary transmission spectra  (see e. g. references in Section~\ref{intro}). However, the gradual change of the Rayleigh slope in the visible and its flattening in the infrared was only recently shown \citep{YB_2016} numerically for Jovian-type exoplanets. Although \citet{YB_LK_2014} and \citet{YB_2016} point out that the effects of refraction on transmission spectra are similar to clouds, we have now demonstrated analytically in Sections~\ref{solution} that all these ``surfaces" cause these same first-order effects.

\begin{figure}
\includegraphics[scale=0.50]{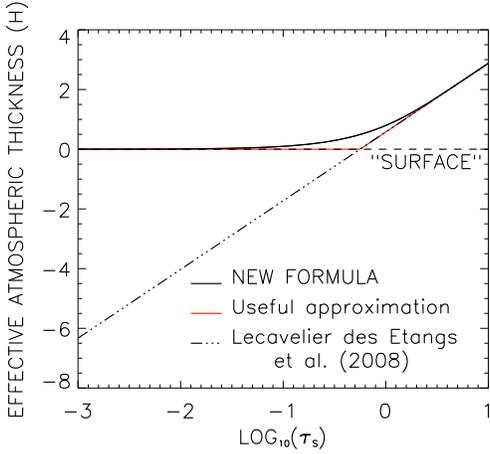}
\caption{Effective thickness of the atmosphere (in units of scale height) as a function
of the slant optical depth of the atmosphere at the ``surface" ($\tau_s$) for $q = 1$ (see equations~\ref{colabun2} and~\ref{sldepthscale}) for different solutions: our solution, given by equation~\ref{hinscale} (solid line), \citet{Lecavelier_2008}'s solution (triple-dot-dashed line), and the approximation (red line) outlined in Section~\ref{index}.
\label{fig3}}
\end{figure}

\subsection{Spectral modulation} \label{spectralmod}

Since the effective atmospheric thickness is proportional to the density scale height of the atmosphere, one quantity is particularly helpful in thinking about exoplanetary transmission spectra: the spectral modulation of an exoplanet. The spectral modulation is the difference in effective atmospheric thickness expressed in units of scale height
\begin{equation}
\frac{\Delta h}{H} = \frac{h_1 - h_2}{H} \label{specmod} ,
\end{equation}
where the subscripts~1 and~2 refers to two different wavelengths which are chosen such that $h_1 > h_2$. It is thus useful to also express the effective atmospheric thickness in units of scale height
\begin{equation}
h/H = \frac{1}{q} \left( \gamma + \ln\tau_{s} + E_1(\tau_{s}) \right) \label{hinscale} 
\end{equation}
where $q$ is defined in Section~\ref{opticaldepth} (see equations~\ref{colabun2} and~\ref{sldepthscale}). 

Fig.~\ref{fig3} shows how $(h/H)$ varies with $\tau_s$ for $q = 1$, both for our expression (equation~\ref{hinscale}) and \citet{Lecavelier_2008}'s (or \citealt{dW_S_2013}'s). Above $\tau_s = 4$, there is no meaningful difference between the two solutions. As $\tau_s$ decreases, our solution for $(h/H)$ approaches the ``surface" without ever going below, unlike \citet{Lecavelier_2008}'s solution which reaches the ``surface" when $\tau_s~=~0.561$, and continues below at smaller~$\tau_s$ values. Our exact solution shows that, as expected, the effective radius of an exoplanet without an atmosphere is simply located at the ``surface". It also shows how far above the ``surface" it is located as the ``surface" slant optical thickness of the atmosphere increases. Since Rayleigh scattering cross sections continuously decrease with wavelength, these two solutions illustrate how the signature of Rayleigh scattering changes with the presence of a ``surface" (see also Fig.~\ref{fig6} in Section~\ref{spectrum}, as well as Fig.~1 through~3 in \citealt{YB_2016}).

The significance of the difference between the two solutions is also evident when considering its impact on the spectral modulation of molecular absorption, which we can infer from Fig.~\ref{fig3}. Let us consider a single line transition where $\tau_s = 10$ at the core of the line. In the presence of a ``surface", the spectral modulation between the core and the wing of the line is never greater than about $2.9$, irrespective of the value of $\tau_s$ in the wings of the line. However, it can be much larger with \citet{Lecavelier_2008}'s solution (e.g. about 6.9 with the $\tau_s = 0.01$ part of the wing). Hence, a ``surface" decreases the atmosphere's spectral modulation. 

We can derive this analytically by combining equations~\ref{specmod} and~\ref{hinscale}. It requires an assumption about the nature of the dominant form of opacity (i.e. about the value of $q$) at each of the two wavelengths that we are comparing. When $q$ is the same at both wavelengths, the spectral modulation simplifies to
\begin{equation}
\frac{\Delta h}{H} = \frac{1}{q} \left[ \ln(\tau_{s1}/\tau_{s2}) + E_1(\tau_{s1}) - E_1(\tau_{s2}) \right] \label{sizefeatures}
\end{equation}
where $\tau_{s1} > \tau_{s2}$. Hence, CIA ($q = 2$) create half the spectral modulation as Rayleigh scattering or molecular absorption ($q = 1$). This simplifies further to
\begin{equation}
\left( \frac{\Delta h}{H} \right)_\sigma = \left[ \ln(\sigma_1/\sigma_2) + E_1(\tau_{s1}) - E_1(\tau_{s2}) \right] \label{smrayl}
\end{equation}
for Rayleigh scattering and molecular absorption, and to 
\begin{equation}
\left( \frac{\Delta h}{H} \right)_k = \frac{1}{2} \left[ \ln(k_1/k_2) + E_1(\tau_{s1}) - E_1(\tau_{s2}) \right] \label{smcia}
\end{equation}
for CIA. 

In a clear atmosphere, $E_1(\tau_{s1}) = E_1(\tau_{s2}) = 0$, and the spectral modulation is then given by the first term in equations~\ref{smrayl} or~\ref{smcia}, i.e. a simple natural logarithm of the ratio of mean cross sections (\citealt{Brown_2001}; \citealt{Sing_2016}). For Rayleigh scattering or molecular absorption, $\sigma$ is an average of the cross sections of various species weighted by their respective mole fractions ($\sigma = \sum_i f_i \sigma_i$). If only one chemical species contribute significantly to the slant optical depth in each spectral region, then $\tau_{s1} \approx N_s f_A \sigma_{1A}$, and $\tau_{s2} \approx N_s f_B \sigma_{2B}$, where subscripts $A$ and $B$ refer to different chemical species, and 
\begin{equation}
\frac{\sigma_{1}}{\sigma_{2}} \approx \left( \frac{f_A}{f_B} \right) \left( \frac{\sigma_{1A}}{\sigma_{2B}} \right) .
\end{equation}
The ratio of the mean cross section at two different wavelengths is then determined not only by the ratio of the cross section of each chemical species but also by their relative abundance. In each spectral region, the dominant species is the one which has the largest $f_i \sigma_i$ product. When several species have comparable products, these should be added together. When the dominant species is the same in both spectral regions, $(\sigma_1/\sigma_2)$ is independent of abundances. In atmospheres with many chemical species, molecular bands of different species will partially fill-in each others' minima, potentially resulting in a lower spectral modulation than when the transmission spectrum is dominated by a single species.

In the presence of a ``surface", $E_1(\tau_{s1})~-~E_1(\tau_{s2})$ is negative, and the spectral modulation is smaller than that of a clear atmosphere. Comparing the observed spectral modulation to that of the clear case may yield clues to the presence of a ``surface". Recent work have indeed tried to infer the cloudiness of hot Jupiters (\citealt{Sing_2016}; \citealt{Stevenson_2016}; \citealt{Iyer_2016}; \citealt{Heng_2016}) through studies of the spectral modulations of water features, as well as that of potassium and sodium lines. Their results hinge upon getting the proper estimate of the density scale height to translate an observed spectral flux difference into a spectral modulation, and comparing it to that of the corresponding species in a clear isothermal atmosphere. 

The theoretical contrast ratio between the ``surface" case and the clear case, or ``surface"-to-clear spectral modulation ratio ($R_{sc}$), is given by
\begin{equation}
R_{sc} = \frac{(\Delta h/H)}{(\Delta h/H)_{clear}} =  1 + \left( \frac{E_1(\tau_{s1}) - E_1(\tau_{s2})}{\ln(\tau_{s1}/\tau_{s2})} \right) \label{rsc}
\end{equation}
and $R_{sc} \le 1$. In the more general case when one compares two spectral regions each with different $q$, it is necessary to go back to the more fundamental expressions (equations~\ref{specmod} and~\ref{hinscale}) to determine $(\Delta h/H)$ and $(\Delta h/H)_{clear}$. As different spectral regions probe different atmospheric depths, one spectral region may be affected by the ``surface" while another is not, depending on the location of the ``surface". Equation~\ref{rsc} shows this because $R_{sc}$ is dependent on $\tau_s$ which combines the opacity of the gas with the location of the ``surface" via the number density of the atmosphere a the ``surface". 

\begin{figure}
\includegraphics[scale=0.5]{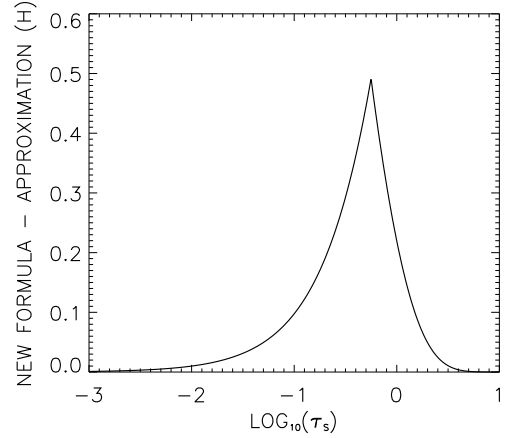}
\caption{Amount by which the effective thickness of the atmosphere (in units of scale height) is underestimated for $q = 1$ (see equations~\ref{colabun2} and~\ref{sldepthscale}) with the approximation outlined in Section~\ref{index} as a function of the ``surface" slant optical depth.
\label{fig4}}
\end{figure}

\subsection{``Surface" cross section}\label{index}

Although useful for computing transmission spectra of isothermal well-mixed atmospheres, equation~\ref{hinscale} does not provide a quick understanding of which spectral features rise above the continuum created by a ``surface", because exponential integrals fall outside the realm of our intuition. We can get around this problem with a simple approximation: We first characterize the location of the ``surface" in term of an equivalent mean atmospheric cross section, or ``surface" cross section ($\sigma_s$ for $q = 1$) within the context of a clear atmosphere, and then define this ``surface" cross section to be the minimum mean cross section which can be observed in the atmosphere. 

To determine this ``surface" cross section, we ask the question: `What mean atmospheric cross section in a clear atmosphere produces an effective atmospheric thickness at the ``surface"?'. It is mathematically equivalent to setting $\tau_{eff} = \tau_s$, which in a clear atmosphere becomes
\begin{equation}
e^{-\gamma} = N_s \sigma_s
\end{equation}
and gives
\begin{equation}
\sigma_s = \frac{e^{-\gamma}}{N_s} = \frac{e^{-\gamma}}{\sqrt{2\pi b_s H} n_s}
\end{equation}
in our non-refractive treatment. The number density at the ``surface" is converted into a mean cross section which is constant at all wavelengths, and which depends on the properties of the exoplanet through its radius ($b_s$) and its scale height. 

\begin{figure*}
\includegraphics[scale=0.70]{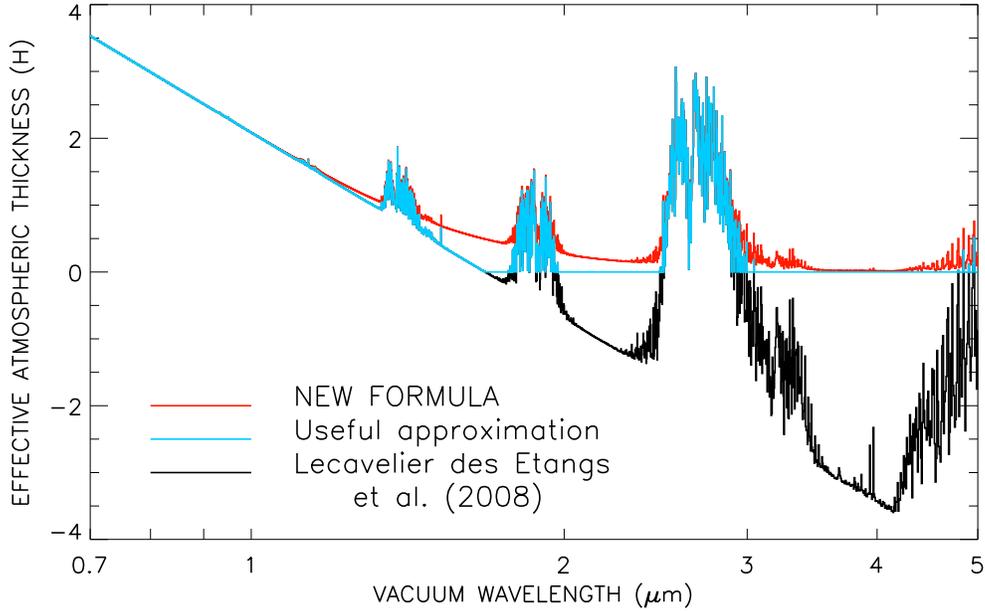}
\caption{Effective atmospheric thickness (expressed in atmospheric scale height) with respect to a reference altitude located at a pressure of 8.1~bar for a H$_2$-He Jovian planet with a homogeneous mole fraction of 10$^{-8}$ of water and a 600~K isothermal profile (see \citealt{YB_2016} for more details). Spectra are computed from the slant optical depth along an unrefracted ray which grazes the reference altitude for three different cases: a clear atmosphere (black) using the formalism of \citet{Lecavelier_2008}, and an atmosphere with a ``surface" at the reference altitude using either the new formula embodied by equation~\ref{hinscale} with $q = 1$ (red), or that of the approximation (light blue) described in Section~\ref{index}. Only opacities from H$_2$-He Rayleigh scattering and absorption from H$_2$O are considered.}
\label{fig6}
\end{figure*}

We then set to $\sigma_s$ any part of the mean cross section which is less than this ``surface" cross section
thereby hiding spectral features located below this continuum. We can then do a quick ``back-of-the-envelope" calculation of the effective atmospheric thickness, using this modified mean cross section ($\sigma^*$), by using equation~\ref{sizefeatures} for a clear atmosphere
\begin{equation}
(h/H)_\sigma \approx \ln(\sigma^*/\sigma_s) \label{approxs} .
\end{equation}
The red curve in Fig.~\ref{fig3} depicts our approximation, which follows the solution of \citet{Lecavelier_2008} when $\sigma > \sigma_s$, and is zero otherwise. Fig.~\ref{fig4} shows the amount by which our approximation underestimates the effective atmospheric thickness. It is at most $E_1(e^{-\gamma})~\approx~0.492$ when $\tau_s~\approx~0.561$. We can apply the same line of reasoning to quickly determine which CIA features rise above the ``surface" continuum. We use
\begin{equation}
(h/H)_k \approx \frac{1}{2} \ln(k^*/k_s) , \label{approxk}
\end{equation}
instead of equation~\ref{approxs}, where $k^*$ is the modified mean collision-induced cross section with its minimum set at the ``surface" collision-induced cross section ($k_s$), given by
\begin{equation}
k_s = \frac{e^{-\gamma}}{\sqrt{\pi b_s H} n_s^2} .
\end{equation}
We then determine the spectral modulation between any two spectral regions with equation~\ref{specmod}. The ``surface" decreases the spectral modulation only if $\sigma_s > \sigma$ or $k_s > k$ in one of the two spectral regions.

Can we use this approximation to quickly determine the location of a ``surface" from observations? If the observed spectral modulation is due to a single species and is less than that expected in a clear atmosphere, we can relate the observed spectral modulation, $(\Delta h/H)_{obs}$,  with respect to a local maximum in its cross section ($\sigma_i$) by
\begin{equation}
\left( \frac{\Delta h}{H} \right)_{obs} = \ln(f_i \sigma_i/\sigma_s) . \label{abuneffect}
\end{equation}
Unfortunately, the ``surface" location can not be unambiguously determined because the mole fraction ($f_i$) of the species is unknown. Indeed, equation~\ref{abuneffect} shows that the spectral modulation of a species increases with abundance in the presence of a ``surface", as demonstrated numerically by \citet{YB_2016} for a refractive boundary. The only way to break this degeneracy are two-fold: either the spectral signature is that of a major species (i.e. $f_i \sim 1$) and we can determine the location of the ``surface", or we can predict the location of the ``surface" and then determine the abundance of the minor species. Given that refractive boundaries are dependent on the bulk properties of the exoplanetary atmosphere as well as the planet-star geometry, quantities which are retrieved during the analysis of transit data, location of refractive boundaries can be predicted in principle and used to place a lower limit on the abundance of detected minor species \citep{YB_2016}. 

\subsection{Transmission spectrum versus formalism}\label{spectrum}

To illustrate the impact of the choice of formalism on a computed transmission spectrum, we apply the following formalisms on simulations by \citet{YB_2016}: the formalism of \citet{Lecavelier_2008}, the exact solution embodied by equation~\ref{hinscale} with $q = 1$, and the approximation discussed in Section~\ref{index}. The chosen atmospheric model is that of a Jovian-like planet with an H$_2$-He atmosphere identical to Jupiter's, a 600~K isothermal profile, and a constant H$_2$O abundance with altitude with a mole fraction of 10$^{-8}$. We choose the slant optical depth computed for an unrefracted ray which grazes a reference altitude located at a pressure of 8.1~bar, and which includes only opacities from Rayleigh scattering and water absorption (see \citealt{YB_2016} for more details of the computation). The resulting transmission spectra are displayed in Fig.~\ref{fig6} from 0.7 to 5.0~\micron~in 4~cm$^{-1}$ wide bins.

The largest difference between the transmission spectra occurs between the clear case of \citet{Lecavelier_2008} and the two cases with a ``surface" at the reference altitude. As explained in Section~\ref{intro}, and also illustrated in Fig.~\ref{fig3}, the spectrum computed with the formalism of \citet{Lecavelier_2008} does not treat ``surfaces" and the effective radius of an exoplanet lies below the reference altitude when the slant optical depth is less than 0.561 at this altitude. The presence of a ``surface" decreases the spectral modulation as no spectral feature is located below the ``surface", both in our exact solution and our approximation.

Although the effective radius differs by less than half a scale height, as predicted from Fig.~\ref{fig4}, between our exact solution and our approximation, there are a few obvious differences in the shape of the two spectra. The most obvious difference is that our approximation produces a sharp change in the Rayleigh slope, similar to that produced by the analytical model of \citet{Sing_2015} (see their Fig.~11 and~12), rather than the gradual change observed in our exact solution, also observed in the simulations of \citet{YB_2016} (see his Fig.~1 through~3). The second difference is that spectral features located below the reference altitude in the clear case are actually preserved but with a much reduced spectral modulation by our exact solution, as predicted by the extra multiplicative factor $(1 - e^{-\tau_s})$ appearing in Eq.~\ref{firstvarwave}. However, in our approximation, they disappear altogether and this results in a flat continuum. Both the gradual change of the Rayleigh slope and the dramatic reduction of the spectral modulation of optically thin features observed in our exact solution were also remarked upon by \citet{YB_2016} in his numerical simulations for refractive atmospheres.

\subsection{Generalized vertical integration scheme}\label{integration}

Although applicable only to an homogeneous isothermal atmosphere, our analytical formalism can be generalized to any atmosphere. Indeed, equation~\ref{results} suggests a way to compute the effective radius from a set of numerically-computed slant optical depths ($\tau_l$) at various impact parameters ($b_l$). To achieve this, we first divide the atmosphere into $M$ layers, so that 
\begin{equation}
R_{eff}^2 = b_s^2 + 2 \sum_{l = 0}^{M-1} \int_{b_l}^{b_{l+1}} (1 - e^{-\tau}) bdb ,
\end{equation}
where $b_0 = b_s$ and $b_M = b_{top}$. We can choose the impact parameter bounding the various layers ($b_l$) so that the slant optical depth scale height is roughly constant in each layer. In this case, it is given by
\begin{equation}
H_{\tau l} = - \frac{(b_{l+1} - b_l)}{\ln(\tau_{l+1}/\tau_l)} 
\end{equation}
where the indices $l+1$ and $l$ refer to the upper and the lower boundaries of each layer, respectively. We need make no assumption about the nature of the opacity as we derive the slant optical depth scale height directly from the results of numerical integration, which can combine arbitrary sources of opacities.

Instead of using $\tau_s$ as our reference slant optical depth for the change of variable (see equation~\ref{varchange}), we use the average of the slant optical depth at the boundaries of each layer ($\tau_{0l}$), or simply
\begin{equation}
\tau_{0l} = (\tau_l + \tau_{l+1})/2 .
\end{equation}
The change of variable then becomes
\begin{equation}
b = b_{0l} - H_{\tau l} \ln(\tau/\tau_{0l})
\end{equation}
where 
\begin{equation}
b_{0l} = b_l - H_{\tau l} \ln(\tau_{0l}/\tau_l) 
\end{equation}
is the impact parameter whose slant optical depth equals~$\tau_{0l}$. The effective radius then becomes
\begin{multline}
R_{eff}^2 = b_s^2 + 2 \sum_{l = 0}^{M-1} H_{\tau l} b_{0l} \int_{\tau_{l+1}}^{\tau_l} (1 - e^{-\tau}) \frac{d\tau}{\tau} \\
                - 2 \sum_{l = 0}^{M-1} H_{\tau l}^2 \int_{\tau_{l+1}}^{\tau_l} (1 - e^{-\tau}) \ln(\tau/\tau_{0l}) \frac{d\tau}{\tau}  \label{genint} .
\end{multline}

One advantage to choosing $\tau_{0l}$ as a reference, is that $\ln(\tau/\tau_{0l})$ is positive over half of the range of slant optical depth within a layer, and negative over the other half, so that the integral in the third term of equation~\ref{genint} is close to zero. Combined with the fact that the factor in front of the third term is a factor of $(b_{0l}/H_{\tau l})$ smaller than that in the second term, and that $(b_{0l}/H_{\tau l}) \gg 1$, the third term is completely negligible. We can then use equation~\ref{results} to solve the integral in the second term, and obtain
\begin{align}
R_{eff}^2 & \approx b_s^2 + 2 \sum_{l = 0}^{M-1} H_{\tau l} b_{0l} \left[ \ln(\tau_l/\tau_{l+1}) + E_1(\tau_{l}) - E_1(\tau_{l+1}) \right] \notag \\
  & \approx b_s^2 + 2 \sum_{l = 0}^{M-1} (b_{l+1} - b_l) b_{0l} \left( 1 - \left<  e^{-\tau} \right>_l \right) 
\end{align}
where 
\begin{equation}
\left<  e^{-\tau} \right>_l = \frac{E_1(\tau_{l}) - E_1(\tau_{l+1})}{\ln(\tau_{l+1}/\tau_{l})} 
\end{equation}
is the atmospheric transmission averaged over the $l^{th}$ atmospheric layer. The effective atmospheric thickness is then given by 
  \begin{align}
 h & \approx \sum_{l = 0}^{M-1} H_{\tau l} \left( \frac{b_{0l}}{b_s} \right) \left[ \ln(\tau_l/\tau_{l+1}) + E_1(\tau_{l}) - E_1(\tau_{l+1}) \right] \notag \\
 & \approx \sum_{l = 0}^{M-1} (b_{l+1} - b_l) \left( \frac{b_{0l}}{b_s} \right) \left( 1 - \left<  e^{-\tau} \right>_l \right)
  \end{align}
where $(b_{0l}/b_s)$ is generally close to but larger than one. 

Our generalized vertical integration scheme is applicable to arbitrary changes of the number density with impact parameter, whether it is due to a change in the mixture of types of opacity or change in density scale height. Since it only requires a series of slant optical depths computed at various impact parameters, the complexity of the physics which is included in the transmission spectrum (e.g. changing gravity with altitude, arbitrary temperature profile, collision-induced absorption, etc.) is only limited by the complexity of the radiative transfer code which computes the slant optical depths, and can be used for refractive atmospheres as long as the optical depths and the impact parameters are computed in an appropriate manner.

\begin{figure}
\includegraphics[scale=0.5]{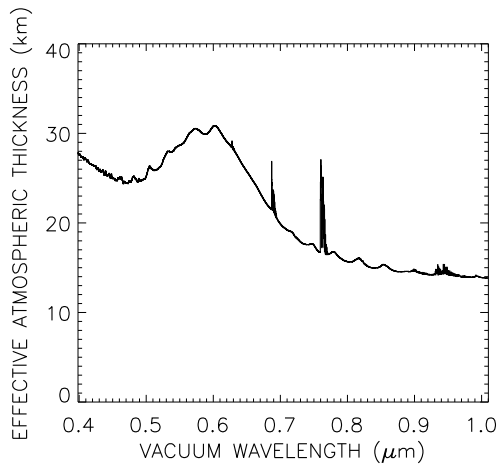}
\caption{Effective atmospheric thickness of Earth viewed as a transiting exoplanet obtained by combining results of simulations from \citet{YB_LK_2013} with the generalized vertical integration scheme of Section~\ref{integration}.
\label{fig7}}
\end{figure}

\begin{figure}
\includegraphics[scale=0.5]{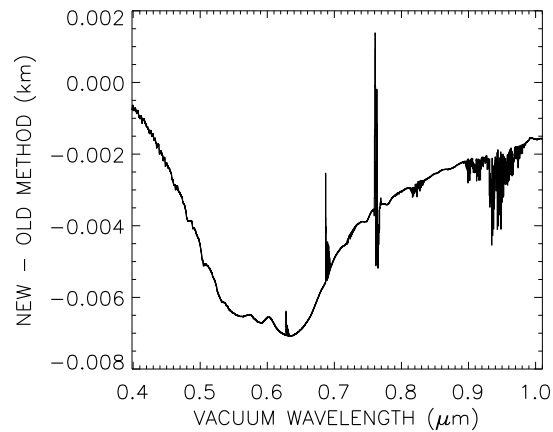}
\caption{Difference in the Earth's effective atmospheric thickness between the vertical integration scheme presented in Section~\ref{integration} (NEW) and the simpler integration scheme embodied by equation~4a (OLD) in \citet{YB_LK_2013}.
\label{fig8}}
\end{figure}

To show this, we use the optical depth as a function of impact parameter from the simulations which include refraction for Earth viewed as an exoplanet of \citet{YB_LK_2013}, and compare the effective atmospheric thickness computed with our new method to that obtained with an older method, described by equation~4a in \citet{YB_LK_2013}. The older method assumes that the transmission of the atmosphere within each atmospheric layer is constant and is simply the average of the values at its upper and lower boundaries. In these simulations, the atmosphere is divided into 80~layers of equal thickness from the critical refractive boundary (12.75~km) to 100~km altitude. Fig.~\ref{fig7} shows the transmission spectrum from 0.4 to 1.0~\micron~in 4~cm$^{-1}$~wide bins obtained with our new method, and Fig.~\ref{fig8} shows the difference between the two methods. The fact that the difference between the two methods is very small at this altitude sampling confirms not only the validity and universality of our new vertical integration scheme, but also validates the underlying formalism on which it is based. Of course, it also suggests that other simpler and faster integration methods can be used to obtain quite accurate results provided the altitude sampling of the simulations is high enough. Our integration scheme can thus also be used to test the accuracy of simpler methods for a given altitude sampling.

\section{Conclusions}

We have derived analytical expressions to compute the effective radius, the effective atmospheric thickness, and the spectral modulation of transiting exoplanets for any source of opacity (Rayleigh scattering, molecular absorption, and collision-induced absorption). Our analytical expressions assume that the atmosphere has a constant density scale height, and ignore stellar limb darkening and second-order refractive effects, just as the analytical expressions by \citet{Lecavelier_2008} and \citet{dW_S_2013} do. However, unlike those previous expressions, our analytical expressions include the effects of a sharp change in flux (``surface") which exists on all planets and is due to an actual surface, an optically thick cloud deck, or a refractive boundary.

We find that the spectral modulation of an exoplanet depends on the nature of the opacity source via its effect on the slant optical depth scale height of the atmosphere. We thus confirm that collision-induced absorption have half the effective scale height as Rayleigh scattering \citep{dW_S_2013}. More importantly, we also find that the effective radius of an exoplanet is located at a slant optical depth which varies with the location of the ``surface". For a clear atmosphere, our solution matches that of \citet{Lecavelier_2008} and \citet{dW_S_2013}. However, as the number density, or the slant optical depth, at the ``surface" decreases, the effective radius of the exoplanet asymptotically approaches the ``surface" without going below, unlike \citet{Lecavelier_2008} and \citet{dW_S_2013}'s solutions. 

Our analytical formalism explains many of the effects of clouds and refraction on transmission spectra that have been previously discussed through numerical simulations. It explains the changing slope of the Rayleigh scattering feature with wavelength first noted by \citet{YB_2016} for refraction. It also explains why clouds and refraction decrease the contrast of spectral features (e. g. \citealt{B_S_2012}; \citealt{H_B_2012}; \citealt{B_S_2013}; \citealt{YB_LK_2014}) and that the resulting spectral modulation increases with the abundance of chemical species \citep{YB_2016}. Our formalism also shows that these effects are a common signature of all ``surfaces", which create a sharp boundary below which stellar radiation is cut-off.

We have also introduced the concept of a ``surface" cross section which is the minimum mean atmospheric cross section of a homogeneous atmosphere which can be observed given the location of the ``surface". This allows us to quickly determine the expected spectral modulation of a species, with less than 0.5 scale height error, given the species's abundance and its peak in cross section. \citet{Heng_2016} recently proposed several indices linked to the spectral modulation of sodium, potassium, and water features to quantify the cloudiness of exoplanets. However, he also points out that they lack universality because each of them are wavelength-specific, and may be affected differently by clouds, presumably because of opacity differences with wavelength. Our concept of a ``surface" cross section is more general precisely because it is wavelength-independent. 

Although much of the discussions in this paper assumes a constant slant optical depth scale height atmosphere, we also developed a numerical recipe to generalize our formalism to atmospheres with non-constant scale heights, as well as those with any mixture of opacity sources. Our general formalism thus allows one to compute the effective radius and the effective atmospheric thickness of an exoplanet with a ``surface", ignoring the effects of stellar limb darkening, from a set of slant optical depths computed numerically at various impact parameters. Refractive effects are automatically included if the slant optical depths and the impact parameters are computed in an appropriate manner. Given that all exoplanets have ``surfaces", our formalism is a crucial step toward the understanding and interpretation of exoplanetary transmission spectra. 

\section*{Acknowledgements}

This research was supported by the appointment of Yan B\'etr\'emieux to the NASA Postdoctoral Program (NPP) at the Jet Propulsion Laboratory, California Institute of Technology. The NPP is administered by Universities Space Research Association (USRA) under contract with NASA. Partial support of this work was also provided by the JPL Exoplanet Science Initiative. \textcopyright~2016. California Institute of Technology. Government sponsorship acknowledged.


\label{lastpage}

\end{document}